\documentclass[a4paper,UKenglish,cleveref,autoref]{lipics-v2021}

\usepackage{import}
\usepackage{amsmath}
\usepackage{amssymb}
\usepackage{listings}
\definecolor{dkviolet}{rgb}{.5,0,.5}
\definecolor{dkblue}{rgb}{0,0,.5}
\definecolor{dkgreen}{rgb}{0,.5,0}
\definecolor{dkred}{rgb}{.5,0,0}
\definecolor{ltblue}{rgb}{0,.4,.7}
\usepackage{lstcoq}
\usepackage{cleveref}
\crefname{figure}{Figure}{figures}
\Crefname{figure}{Figure}{Figures}

\usepackage{proof-dashed}
\usepackage{xypic}
\newcommand*{\rulebreak}{\\[1em]}

\newcommand{\NN}{\ensuremath{\mathbb N}}

\title{Now It Compiles!}
\subtitle{Certified Automatic Repair of Uncompilable Protocols}

\titlerunning{Now It Compiles!}

\author{Luís Cruz-Filipe}{Department of Mathematics and Computer Science, University of Southern Denmark}{lcfilipe@gmail.com}{https://orcid.org/0000-0002-7866-7484}{}

\author{Fabrizio Montesi}{Department of Mathematics and Computer Science, University of Southern Denmark}{fmontesi@imada.sdu.dk}{https://orcid.org/0000-0003-4666-901X}{}

\authorrunning{L. Cruz-Filipe and F. Montesi}

\Copyright{Luís Cruz-Filipe and Fabrizio Montesi}



\begin{document}

\maketitle

\begin{abstract}
Choreographic programming is a paradigm where developers write the global specification (called choreography) of a communicating system, and then a correct-by-construction distributed implementation is compiled automatically.
Unfortunately, it is possible to write choreographies that cannot be compiled, because of issues related to an agreement property known as knowledge of choice.
This forces programmers to reason manually about implementation details that may be orthogonal to the protocol that they are writing.

Amendment is an automatic procedure for repairing uncompilable choreographies.
We present a formalisation of amendment from the literature, built upon an existing formalisation of choreographic programming.
However, in the process of formalising the expected properties of this procedure, we discovered a subtle counterexample that invalidates the original published and peer-reviewed pen-and-paper theory.
We discuss how using a theorem prover led us to both finding the issue, and stating and proving a correct formulation of the properties of amendment.
\end{abstract}

\section{Introduction}
\label{sec:intro}

Programming correct implementations of protocols for communicating systems is challenging, because it requires writing a correct program for each participant that performs the right send and receive actions at the right times~\cite{LLLG16}.
\emph{Choreographic programming}~\cite{M13p} is an emerging paradigm that offers a direct solution: protocols are written in a ``choreographic'' programming language, and then automatically compiled to correct implementations by means of an operation known as \emph{Endpoint Projection} (EPP or projection for short)~\cite{CM13,DGGLM17,GMP20,HG22,JB22,LN15,LH17}.

Choreographic languages are inspired by the Alice and Bob notation of security protocol~\cite{NS78}, in the sense that they offer primitives for expressing communications between different processes. Implementations are usually modelled in terms of a process calculus.
Besides being simple, choreographic programming is interesting because it typically includes strong theoretical guarantees, most notably deadlock-freedom and an operational correspondence between choreographies and the (models of the) generated distributed implementations.

Not all choreographies can be compiled (or \emph{projected}) to a distributed implementation due to a problem known as ``knowledge of choice''~\cite{CDP11}. Consider the following choreography for a simple purchase scenario (this example also anticipates some of our syntax).
\begin{lstlisting}[caption={An unprojectable choreography.},label={eq:bs-unprojectable}]
buyer.offer --> seller.x;
If seller.acceptable(x) Then seller.product --> buyer.y; End
                          Else End
\end{lstlisting}
This choreography reads: a \lstinline+buyer+ communicates their offer for the purchase of a product to a \lstinline+seller+, who stores the offer in their local variable \lstinline+x+; the \lstinline+seller+ then checks whether the offer is acceptable, and in the affirmative case sends the \lstinline+product+ to \coc{buyer}.
This choreography cannot be projected to a behaviourally-equivalent implementation, because \lstinline+buyer+ has to behave differently in the two branches of the conditional. However, this conditional is evaluated by \lstinline+seller+, and \coc{buyer} has no way of discerning which branch gets chosen.

Choreographies are typically made projectable by adding selections, i.e., communications of constants called \emph{selection labels}.\footnote{Selections are essentially the choreographic version of branch selections in session types, or the additive connectives in linear logic~\cite{CLMSW16,HYC16}.}
A projectable version of \cref{eq:bs-unprojectable} looks as follows.
\begin{lstlisting}[caption={A projectable choreography.},label={eq:bs-projectable}]
buyer.offer --> seller.x;
If seller.acceptable(x) Then seller --> buyer[left]; seller.product --> buyer.y; End
                         Else seller --> buyer[right]; End
\end{lstlisting}
This choreography differs from the previous one by the presence of a selection in each branch of the conditional. Specifically, if \lstinline+seller+ chooses the \coc{Then} branch, they now communicate the label \lstinline+left+ to \lstinline+buyer+. Otherwise, if the \coc{Else} branch is chosen, the label \lstinline+right+ is communicated instead.
The key idea is that now the implementation generated for \lstinline+buyer+ can read the label received by \lstinline+seller+ and know which branch of the conditional should be executed. Since labels are constants, compilation can statically verify that \lstinline+buyer+ receives different labels for the different branches, and therefore has ``knowledge of choice''.

Projection can be smart about knowledge of choice, allowing selections to be kept to a minimum~\cite{CHY12}. A process only needs to know which branch of a conditional has been chosen if its behaviour depends on that choice; if the process has to perform the same actions in both branches of a conditional, then this knowledge is irrelevant to it. Knowledge of choice can also be propagated: if a process \coc{q} knows of a choice performed by another process \coc{p}, then either process can forward this information to any other process that needs it.

\subparagraph{Amendment.}
Previous work investigated how unprojectable choreographies can be automatically transformed into projectable ones. Such a transformation is called \emph{amendment}~\cite{CM20,LMZ13} or repair~\cite{BB16,FL17}.
For example, applying the amendment procedure from~\cite{CM20} to the choreography in \cref{eq:bs-unprojectable} returns the choreography in \cref{eq:bs-projectable} (up to minor differences in notation).

Amendment is interesting for (at least) two reasons. On a practical level, it can suggest valid selection strategies to developers to make their choreographies  executable---or even do it automatically, so that they do not have to worry about knowledge of choice. On a theoretical level, it allows porting completeness properties of the set of all choreographies to the set of projectable choreographies.

An example of the latter occurs in the study of \emph{Core Choreographies} (CC), a minimalistic theory of choreographic programming~\cite{CM20}, where we show that the set of projectable choreographies in CC is Turing-complete in two steps.
First, we show that CC is Turing-complete, ignoring the question of projectability (the choreographies constructed in the proof are clearly not projectable); then, we define an amendment procedure and prove an operational correspondence between choreographies and their amendments.
As a consequence, the subset of projectable choreographies is also Turing-complete.
A similar argument, using the operational correspondence result between projectable choreographies and their implementations, shows that the process calculus used (\emph{Stateful Processes}, or SP) is Turing-complete.

\subparagraph{The problem.}
Our original objective was to formalise amendment and its properties from~\cite{CM20} in the Coq theorem prover, building upon our previous formalisation of CC~\cite{CMP21a} and its accompanying notion of projection \cite{CMP21b}.
That formalisation uses a variation of CC based on the theory from~\cite{M22}, which we found more amenable to formalisation.
Unfortunately, after formalising the definition of amendment, our attempt to prove its operational correspondence result failed. An inspection of the state of the failed proof quickly led us to a counterexample.

The incorrectness of the original statement jeopardises the subsequent developments that rely on it, in particular Turing completeness of the set of projectable choreographies and of SP.
These results were instrumental in substantiating the claim that CC is a ``good'' minimalistic model for choreographic programming.
This finding pointed us towards a more ambitious goal: reformulate the operational correspondence for amendment such that it is correct, and still powerful enough to obtain the aforementioned consequences.

\subparagraph{Contribution.}
To the best of our knowledge, this is the first time that choreography amendment has been formalised.
We state and prove a relaxed version of the operational correspondence between choreographies and their amendments in the Coq theorem prover, thus increasing confidence in its correctness.
We discuss how working with an interactive theorem prover was instrumental to identifying counterexamples that guided us towards this new, correct formulation that considers all corner cases.
We then use our result to formalise the proofs of Turing completeness of projectable choreographies and SP from~\cite{CM20}, which were not included in~\cite{CMP21a}.

\subparagraph{Structure of the paper.}
We present the relevant background on CC and its formalisation in \cref{sec:background}. \Cref{sec:amend} presents the definition of amendment, its formalisation, and discusses and corrects the operational correspondence result from~\cite{CM20}. \Cref{sec:turing} shows that the revised semantic property is still strong enough to derive the Turing completeness results in that work.
We discuss related work in \cref{sec:related} and conclude in \cref{sec:conclusion}.

Our exposition assumes some familiarity with interactive theorem proving.
We include some Coq code in the article, but the work is intended to be accessible to non-Coq experts.

\section{Background}
\label{sec:background}

We summarise the latest version of the Coq formalisation of CC~\cite{CMP22}. For simplicity, we omit two ingredients that are immaterial for our work: the fact that the language is parameterised on a signature, and the fact that communications have annotations (these are meant to include information relevant for future implementations in actual programming languages).
This allows us to omit some subterms that play no role in the development of amendment.

In our presentation, we use Coq notation with some simplifications for enhanced readability: choreography and process terms are written overloading dots (this is not allowed by the Coq notation mechanism), and inductive definitions and inference rules are given with the usual mathematical notation.

\subsection{Core Choreographies}

We start by giving an overview of Core Choreographies (CC) together with its formalisation in Coq~\cite{CMP21a}.

\subparagraph{Syntax.}
The syntax of CC is given by the following grammar.
\begin{lstlisting}
C ::= $\eta$; C $\mid$ If p.b Then C1 Else C2 $\mid$ Call X $\mid$ RT_Call X ps C $\mid$ End
$\eta$ ::= p.e --> q.x $\mid$ p --> q[l]
\end{lstlisting}
A choreography \lstinline+C+ can be either: a communication $\eta$ followed by a continuation (\lstinline+$\eta$; C+); a conditional \lstinline+If p.b Then C1 Else C2+, where the process \lstinline+p+ evaluates the boolean expression \lstinline+b+ to choose between the branches \lstinline+C1+ and \lstinline+C2+; a procedure call \lstinline+Call X+, where \lstinline+X+ is the name of the procedure being invoked; a runtime term \coc{RT_Call X ps C};\footnote{Runtime terms are needed for technical reasons in the definition of the semantics of choreographies~\cite{CMP21a}. These aspects are irrelevant for the present development.} or the terminated choreography \lstinline+End+.
A communication $\eta$ can be: a value communication \lstinline+p.e --> q.x+, read ``process \lstinline+p+ evaluates expression \lstinline+e+ locally and sends the result to process \lstinline+q+, which stores it in its locally in \lstinline+x+''; or a selection \lstinline+p --> q[l]+, where the label \lstinline+l+ can be either \lstinline+left+ or \lstinline+right+, read ``\lstinline+p+ sends label \lstinline+l+ to \lstinline+q+''.

\begin{table*}
\begin{tabular}{l|l|l}
\textbf{Type} & \textbf{Variable} & \textbf{Description}
\\
\hline
\lstinline+Choreography+ & \lstinline+C+ & Choreographies
\\
\lstinline+Pid+ & \lstinline+p+, \lstinline+q+, \lstinline+r+, \lstinline+s+ & Process names (identifiers)
\\
\lstinline+list Pid+ & \lstinline+ps+ & List of process names
\\
\lstinline+Var+ & \lstinline+x+, \lstinline+y+, \lstinline+z+ & Variable names
\\
\lstinline+Val+ & \lstinline+v+ & Values
\\
\lstinline+Expr+ & \lstinline+e+ & Expressions (evaluate to values)
\\
\lstinline+BExpr+ & \lstinline+b+ & Boolean expressions (evaluate to Booleans)
\\
\lstinline+Label+ & \lstinline+l+ & Labels (\lstinline+left+ and \lstinline+right+)
\\
\lstinline+RecVar+ & \lstinline+X+ & Procedure names (or recursive variables)
\\
\lstinline+DefSet+ & \lstinline+D+ & Sets of procedure definitions in CC
\\
\lstinline+State+ & \lstinline+s+ & Maps from variables to values
\\
\lstinline+Configuration+ & \lstinline+c+ & Choreographic programs equipped with states
\\
\lstinline+TransitionLabel+ & \lstinline+t+ & Transition labels
\\
\lstinline+list TransitionLabel+ & \lstinline+tl+ & Lists of transition labels
\\
\lstinline+Behaviour+ & \lstinline+B+ & Behaviours
\\
\lstinline+option Behaviour+ & \lstinline+mB+ & \coc{option} monad for behaviours
\\
\lstinline+Network+ & \lstinline+N+ & Networks
\\
\lstinline+DefSetB+ & \lstinline+D+ & Sets of procedure definitions in SP
\\
\lstinline+Program+ & \lstinline+P+ & Choreographies/networks with procedure definitions
\end{tabular}

\caption{Summary of types in the original Coq formalisation~\cite{CMP21a,CMP21b}.}
\label{table:types}
\end{table*}
Choreographies are formalised in Coq as an inductive type called \lstinline+Choreography+.
\Cref{table:types} summarises the Coq types used in this paper and our conventions for ranging over their elements.

Executing a choreography requires knowing the definitions of the choreographies associated to the procedures that can be invoked, as well as the processes involved in those procedures.
A set of procedure definitions is defined as a mapping from procedure names to pairs of process names and choreographies.
\begin{lstlisting}
Definition DefSet := RecVar -> (list Pid)*Choreography.
\end{lstlisting}
A \emph{choreographic program} is then a pair consisting of a set of procedure definitions and a choreography (which represents the ``main'' or ``running'' choreography).
\begin{lstlisting}
Definition Program := DefSet * Choreography.
\end{lstlisting}
We write \coc{Procedures P} and \coc{Main P} for the two components of \coc{P}.
The set of all processes used by a program \coc{P} is defined as \coc{CCP_pn P}.

It is standard practice to assume some well-formedness conditions about choreographies, e.g., that no process communicates with itself.
Choreographic programs have additional well-formedness conditions that must hold for all procedures that can be reached at runtime.
This notion is not decidable in general, but it becomes so in the practical case of programs that only use a finite number of procedures.
We return to this aspect at the end of \cref{sec:amend-syntax}, where it becomes relevant.

\begin{example}
The choreographies in \cref{eq:bs-unprojectable,eq:bs-projectable} are well-formed.
\end{example}

\subparagraph{Semantics.}
The intuitive system assumptions in CC are that: processes run independently of each other (concurrently) and possess local stores (associating their variables to values); communications are synchronous; and the network is reliable (messages are not lost nor duplicated, and they are delivered in the right order between any two processes). These assumptions are imported from process calculi, where they are quite standard.

\begin{example}\label{ex:ooe}
Since processes run concurrently, it is possible to express choreographies with concurrent behaviour.
Consider the following simplification of the factory example in~\cite{M22}.
\begin{lstlisting}[caption={Parallel orders.},label={eq:ooe-example}]
o.order --> p.x; o'.order' --> p'.y; End
\end{lstlisting}
In \cref{eq:ooe-example}, two processes \lstinline+o+ and \lstinline+o'+ place their respective orders to two different providers \lstinline+p+ and \lstinline+p'+.
Since all processes are distinct and there is no causal dependency between the two communications, the two communications can in principle be executed in any order.
This gives rise to a notion of out-of-order execution for choreographies.
\end{example}

The semantics of choreographies in~\cite{CMP21a} is given as a labelled transition system on configurations, which consist of a program and a (memory) state. States associate to each process a map from variable names to values, which defines the memory of that process.
\begin{lstlisting}
Definition State := Pid -> Var -> Value
\end{lstlisting}
States come with some notation: \lstinline+s [==] s'+ says that \lstinline+s+ and \lstinline+s'+ are extensionally equal, and \lstinline+s[[p,x => v]]+ is the state obtained from updating \lstinline+s+ with the mapping \lstinline+p,x $\mapsto$ v+.

\begin{figure*}
  $$\begin{array}c
    \infer[\mbox{\lstinline+CC_Com+}]
          {\mbox{\lstinline+(D,p.e --> q.x; C,s) --[TL_Com p v q]--> (D,C,s')+}}
          {\mbox{\lstinline+v := eval e s p+}
            & \mbox{\lstinline+s' [==] s[[q,x => v]]+}}
    \rulebreak
    \infer[\mbox{\lstinline+CC_Sel+}]
          {\mbox{\lstinline+(D,p --> q[l]; C,s) --[TL_Sel p q l]--> (D,C,s')+}}
          {\mbox{\lstinline+s [==] s'+}}
    \rulebreak
    \infer[\mbox{\lstinline+CC_Then+}]
          {\mbox{\lstinline+(D,If p.b Then C1 Else C2,s) --[TL_Tau p]--> (D,C1,s')+}}
          {\mbox{\lstinline+beval b s p = true+}
            & \mbox{\lstinline+s [==] s'+}}
    \rulebreak
     \infer[\mbox{\lstinline+C_Else+}]
           {\mbox{\lstinline+(D,If p.b Then C1 Else C2,s) --[TL_Tau p]--> (D,C2,s')+}}
           {\mbox{\lstinline+beval b s p = false+}
             & \mbox{\lstinline+s [==] s'+}}
    \rulebreak
    \infer[\mbox{\lstinline+CC_Delay_Eta+}]
          {\mbox{\lstinline+(D,$\eta$; C,s) --[t]--> (D,$\eta$; C',s')+}}
          {\mbox{\lstinline+disjoint_eta_rl \\eta t+}
            & \mbox{\lstinline+(D,C,s) --[t]--> (D,C',s')+}}
    \rulebreak
    \infer[\mbox{\lstinline+CC_Delay_Cond+}]
          {\mbox{\lstinline+(D,If p.b Then C1 Else C2,s) --[t]--> (D,If p.b Then C1' Else C2',s')+}}
          {\mbox{\lstinline+disjoint_p_rl p t+}
            &
              \mbox{\lstinline+(D,C1,s) --[t]--> (D,C1',s')+}
            &
              \mbox{\lstinline+(D,C2,s) --[t]--> (D,C2',s')+}
          }
  \end{array}$$
  \caption{Semantics of choreographic configurations (selected rules).}
  \label{fig:cc-semantics}
\end{figure*}
With these concepts in place, we can show some representative transition rules for choreographic configurations in \cref{fig:cc-semantics}.\footnote{In the actual formalisation, the transition relation was defined in two layers for technical reasons. This technicality is immaterial for our development, since our results follow from the rules shown here.}
Transitions have the form \lstinline+(D,C,s) --[t]--> (D,C',s')+, where \lstinline+t+ is a transition label that allows for observing what happened in the transition.

Rule \lstinline+CC_Com+ deals with the execution of a value communication from a process \lstinline+p+ to a process \lstinline+q+: if the expression \lstinline+e+ at \lstinline+p+ can be evaluated to a value \lstinline+v+ (first condition, which uses the auxiliary function \lstinline+eval+), then the communication term is consumed and the state of the receiver is updated such that its receiving variable \lstinline+x+ is now mapped to value \lstinline+v+. The transition label \lstinline+TL_Com p v q+ denotes that \lstinline+p+ has communicated the value \lstinline+v+ to \lstinline+q+, modelling what would be visible on a network.

Rule \lstinline+CC_Sel+ is similar but does not alter the state of the receiver (the role of selections will be clearer when we explain the language for modelling implementations of choreographies). The transition label \lstinline+TL_Sel p q l+ registers the communication of label \lstinline+l+ from \lstinline+p+ to \lstinline+q+.

Rule \lstinline+CC_Then+ deals with the case in which a process \lstinline+p+ can evaluate the guard \lstinline+b+ of a conditional to \lstinline+true+ (using the auxiliary function \lstinline+beval+), proceeding to the then-branch of the conditional. The transition label \lstinline+TL_Tau p+ denotes that process \lstinline+p+ has executed an internal action ($\tau$ is the standard symbol for such actions in process calculi).

Rule \lstinline+CC_Delay_Eta+ deals with out-of-order execution of communications, formalising the reasoning anticipated in \cref{ex:ooe}. Specifically, the continuation of an interaction $\eta$ is allowed to perform a transition (without affecting $\eta$) as long as the transition does not involve any of the processes in $\eta$. The latter condition is checked by the first premise of the rule, \lstinline+disjoint_eta_rl $\eta$ t+, which checks that the processes mentioned in $\eta$ are distinct from those mentioned by the transition label \lstinline+t+.
Rule \coc{CC_Delay_Cond} applies the same reasoning to the out-of-order execution of conditionals.

The reflexive and transitive closure of the transition relation is written \lstinline+--[tl]-->**+, where \lstinline+tl+ is a list of transition labels.

\begin{example}
For any \lstinline+D+ and \lstinline+s+ such that \lstinline+order+ evaluates to \lstinline+v+ at \lstinline+o+ and \lstinline+order'+ evaluates to \lstinline+v'+ at \lstinline+o'+, according to \lstinline+eval+,
$$
\mbox{\lstinline+(D,o.order --> p.x; o'.order' --> p'.y; End,s)+}
\mbox{\lstinline+--[TL_Com o v p;TL_Com o'$\ $v'$\ $p']-->** (D,End,s')+}
$$
and
$$
\mbox{\lstinline+(D,o.order --> p.x; o'.order' --> p'.y; End,s)+}
\mbox{\lstinline+--[TL_Com o'$\ $v'$\ $p';TL_Com o v s]-->** (D,End,s')+}
$$
where \lstinline+s' [==] s[[s1,x => v]][[s2,x => v']]+.
\end{example}

\subsection{Processes}
Implementations of choreographies are modelled in Stateful Processes (SP)~\cite{CMP21b}, a formalised process calculus following~\cite{M22}.
SP follows the standard way of representing systems of communicating processes, where the code of each process is given separately and communication is achieved when processes perform compatible I/O actions.

\subparagraph{Syntax.}
The code of a process is written as a behaviour (\coc{B}), following the grammar below.
\begin{lstlisting}
B ::= p!e; B $\mid$ p?x; B $\mid$ p+l; B $\mid$ p & mB1 // mB2 $\mid$ If b Then B1 Else B2 $\mid$ Call X $\mid$ End
mB ::= None $\mid$ Some B
\end{lstlisting}
These terms are the local counterparts to the choreographic terms of CC.
The first two productions deal with value communication.
Specifically, a send action \lstinline+p!e; B+ sends the result of evaluating \lstinline+e+ to the process \lstinline+p+ and then continues as \lstinline+B+. Dually, a receive action \lstinline+p?x; B+ receives a value from \lstinline+p+, stores it in \lstinline+x+, and then continues as \lstinline+B+.

Selections are implemented by the primitives \lstinline|p+l; B| and \lstinline+p & mB1 // mB2+. The former sends the label \lstinline+l+ to the process \lstinline+p+ and continues as \lstinline+B+.
The latter is a branching term, where \lstinline+mB1+ and \lstinline+mB2+ are the behaviours that the process will execute upon receiving \coc{left} or \coc{right}, respectively.
To cover the case where a process does not offer a behaviour for a specific label, \coc{mB1} and \coc{mB2} have type \coc{option Behaviour}.

Conditionals (\lstinline+If b Then B1 Else B2+), procedure calls (\lstinline+Call X+), and the terminated behaviour (\coc{End}) are standard.

Processes are intended to run together in networks. These are formalised as maps from processes to behaviours.
\begin{lstlisting}
Definition Network := Pid -> Behaviour.
\end{lstlisting}
Networks come with some convenient notation for their construction:
\lstinline+p[B]+ is the network that maps \lstinline+p+ to \lstinline+B+ and all other processes to \lstinline+End+; and \lstinline+N | N'+ is the composition of \lstinline+N+ and \lstinline+N'+. 
In particular, \lstinline+(N | N') p+ returns \lstinline+N p+ if this is different from \lstinline+End+, and \lstinline+N' p+ otherwise.\footnote{This asymmetry does not matter for our results, since we never compose networks that define nonterminated behaviours for the same processes.}

\begin{example}\label{ex:bs-epp}
The following network implements the choreography in \cref{eq:bs-projectable}.
\begin{lstlisting}
buyer[ seller!offer; seller & Some (seller?y; End) // Some End ] |
seller[ buyer?x; If acceptable(x) Then buyer+left; buyer!product; End
                                   Else buyer+right; End ]
\end{lstlisting}
\end{example}

For the semantics of networks, we need two additional ingredients. The network \coc{N \\ p} is obtained from \coc{N} by redefining \lstinline+p+'s behaviour as \coc{End} (\coc{p} is ``removed'' from \coc{N}).
The relation \lstinline+N (==) N'+ holds if the networks \lstinline+N+ and \lstinline+N'+ are extensionally equal.

As in CC, processes in a network can invoke procedures defined in a separate set.
\begin{lstlisting}
Definition DefSetB := RecVar -> Behaviour.
\end{lstlisting}
A \lstinline+Program+ in SP consists of a set of procedure definitions and a network.
\begin{lstlisting}
Definition Program := DefSetB * Network.
\end{lstlisting}
We use \lstinline+D+ to range over elements of \lstinline+DefSetB+ and \lstinline+P+ to range over elements of \lstinline+Program+, as for choreographies (the difference will be clear from the context).

\subparagraph{Semantics.}
The semantics of SP is also given as a labelled transition system on configurations that consist of a program and a memory state, as in CC.
A selection of the transition rules defining this semantics is displayed in \cref{fig:sp-semantics}.

\begin{figure*}
  \begin{eqnarray*}
    &
    \infer[\mbox{\lstinline+SP_Com+}]
          {\mbox{\lstinline+(D,N,s) --[TL_Com p v q]--> (D,N',s')+}}
          {\begin{array}c \mbox{\lstinline+N p = q!e; B+} \\
              \mbox{\lstinline+N q = p?x; B'+} \end{array}
            &
              \mbox{\lstinline+v := eval e s p+} &
              \begin{array}c\mbox{\lstinline+N' (==) N \\ p \\ q | p[B] | q[B']+} \\
              \mbox{\lstinline+s' [==] s[[q,x => v]]+} \end{array}}
    \rulebreak
    &\infer[\mbox{\lstinline+SP_LSel+}]
          {\mbox{\lstinline+(D,N,s) --[TL_Sel p q left]--> (D,N',s')+}}
          {
              \begin{array}c \mbox{\lstinline"N p = q+left; B"} \\
              \mbox{\lstinline+N q = p & Some Bl // mBr+} \end{array}
            & \mbox{\lstinline+N' (==) N \\ p \\ q | p[B] | q[Bl]+}
            & \mbox{\lstinline+s [==] s'+}}
    \rulebreak
    &\infer[\mbox{\lstinline+SP_RSel+}]
          {\mbox{\lstinline+(D,N,s) --[TL_Sel p q right]--> (D,N',s')+}}
          {
            \begin{array}c \mbox{\lstinline"N p = q+right; B"} \\
              \mbox{\lstinline+N q = p & mBl // Some Br+} \end{array}
            & \mbox{\lstinline+N' (==) N \\ p \\ q | p[B] | q[Br]+}
            & \mbox{\lstinline+s [==] s'+}}
    \rulebreak
    &\infer[\mbox{\lstinline+SP_Then+}]
          {\mbox{\lstinline+(D,N,s) --[TL_Tau p]--> (D,N',s')+}}
          {\mbox{\lstinline+N p = If b Then B1 Else B2+} &
              \mbox{\lstinline+beval b s p = true+}
              &
            & \begin{array}c\mbox{\lstinline+N' (==) N \\ p | p[B1]+} \\
                \mbox{\lstinline+s [==] s'+} \end{array}}
  \end{eqnarray*}
  \caption{Semantics of network configurations (selected rules).}
  \label{fig:sp-semantics}
\end{figure*}
Rule \lstinline+SP_Com+ matches a send action at a process \lstinline+p+ with a compatible receive action at another process \lstinline+q+ (conditions \lstinline+N p = q!e; B+ and \lstinline+N q = p?x; B'+).
The resulting network \lstinline+N'+ is obtained from \lstinline+N+ by replacing the behaviours of these processes with their continuations (\coc{N \\ p \\ q | p[B] | q[B']}). The update to the state is handled as in CC.

Rules \lstinline+SP_LSel+ and its dual \lstinline+SP_RSel+ model, respectively, the selection of the left and right branches offered by a branching term, by inspecting the label sent by the sender. Rule \lstinline+SP_Then+ captures the case in which a conditional enters its then-branch.

\subsection{Endpoint Projection (EPP)}
Choreographies are compiled to networks by a procedure called behaviour projection.
This procedure is a partial function, and since all functions in Coq are total it was formalised as the following inductive relation.
\begin{lstlisting}
bproj : DefSet -> Choreography -> Pid -> Behaviour -> Prop
\end{lstlisting}
Term \lstinline+bproj D C p B+, written \coc{[[D,C | p]] == B},  reads ``the projection of \coc{C} on \coc{p} in the context of the set of procedure definitions \coc{D} is \coc{B}''.\footnote{The parameter \coc{D} is used for projecting procedure calls, which is irrelevant for our development.}

\begin{figure*}
  $$\begin{array}c
    \infer[\mbox{\lstinline+bproj_Send+}]
          {\mbox{\lstinline+[[D,p.e --> q.x; C | p]] == q!e; B+}}
          {\mbox{\lstinline+[[D,C | p]] == B+}}
    \qquad
    \infer[\mbox{\lstinline+bproj_Recv+}]
          {\mbox{\lstinline+[[D,p.e --> q.x; C | q]] == p?x; B+}}
          {\mbox{\lstinline+p<>q+}
            & \mbox{\lstinline+[[D,C | q]] == B+}}
    \rulebreak
    \infer[\mbox{\lstinline+bproj_Com+}]
          {\mbox{\lstinline+[[D,p.e --> q.x; C | r]] == B+}}
          {\mbox{\lstinline+p<>r+}
            & \mbox{\lstinline+q<>r+}
            & \mbox{\lstinline+[[D,C | r]] == B+}}
    \rulebreak
    \infer[\mbox{\lstinline+bproj_Pick+}]
          {\mbox{\lstinline"[[D,p --> q[l]; C | p]] == q+l; B"}}
          {\mbox{\lstinline+[[D,C | p]] == B+}}
    \qquad
    \infer[\mbox{\lstinline+bproj_Sel+}]
          {\mbox{\lstinline+[[D,p --> q[l]; C | r]] == B+}}
          {\mbox{\lstinline+p<>r+}
            & \mbox{\lstinline+q<>r+}
            & \mbox{\lstinline+[[D,C | r]] == B+}}
    \rulebreak
    \infer[\mbox{\lstinline+bproj_Left+}]
          {\mbox{\lstinline+[[D,p --> q[left]; C | q]] == p & Some B // None+}}
          {\mbox{\lstinline+p<>q+}
            & \mbox{\lstinline+[[D,C | q]] == B+}}
    \rulebreak
    \infer[\mbox{\lstinline+bproj_Right+}]
          {\mbox{\lstinline+[[D,p --> q[right]; C | q]] == p & None // Some B+}}
          {\mbox{\lstinline+p<>q+}
            & \mbox{\lstinline+[[D,C | q]] == B+}}
    \rulebreak
    \infer[\mbox{\lstinline+bproj_Cond+}]
          {\mbox{\lstinline+[[D,If p.b Then C1 Else C2 | p]] == If b Then B1 Else B2+}}
          {\mbox{\lstinline+[[D,C1 | p]] == B1+}
            & \mbox{\lstinline+[[D,C2 | p]] == B2+}}
    \rulebreak
    \qquad
    \infer[\mbox{\lstinline+bproj_Cond'+}]
          {\mbox{\lstinline+[[D,If p.b Then C1 Else C2 | p]] == B+}}
          {\mbox{\lstinline+p<>r+}
            & \mbox{\lstinline+[[D,C1 | p]] == B1+}
            & \mbox{\lstinline+[[D,C2 | p]] == B2+}
            & \mbox{\lstinline+B1 [V] B2 == B+}}
  \end{array}$$
  \caption{Selected rules for behaviour projection.}
  \label{fig:bproj}
\end{figure*}
Intuitively, behaviour projection is computed by going through the choreography; for each choreographic term, projection constructs the local action that the input process should perform to implement it.
The rules defining \coc{bproj} that are relevant for this work are those that deal with selections and conditionals.
These are shown in \cref{fig:bproj}.

A label selection \coc{p --> q[l]} is projected as either:
(i)~the sending of label \coc{l} to \coc{q} for process \coc{p} (rule \coc{bproj_Pick});
(ii)~the appropriate branching term that receives \coc{l} from \coc{p} for process \coc{q}, where only the branch for \coc{l} offers a behaviour (rules \coc{bproj_Left} and the dual rule \coc{bproj_Right});
or (iii)~no action for any other process (rule \coc{bproj_Sel}).

Similarly, a conditional in a choreography is projected to a conditional for the process that evaluates the guard (rule \coc{bproj_Cond}).
However, projecting conditionals becomes complex when considering the other processes, because this requires dealing with the problem of knowledge of choice discussed in \cref{sec:intro}. This case is handled by rule \coc{bproj_Cond'}, which sets the result of projection to be the ``merging'' of the projections of the two branches, written \coc{B1 [V] B2 == B}, if this is defined.

Intuitively, merging attempts to build a behaviour \coc{B} from two behaviours \coc{B1} and \coc{B2} that have similar structures, but may differ in the labels that they accept in branching terms.
For all terms but branchings, merging requires term equality and then proceeds homomorphically in subterms. This is exemplified by the rules \coc{merge_End}, \coc{merge_Sel}, and \coc{merge_Cond} in \cref{fig:merge}.

The interesting part regards the merging of branching terms, which has a rule for every possible combination. \Cref{fig:merge} shows two representative cases. If two branching terms have branches for different labels, then we obtain a branching term where the two branches are combined as exemplified by rule \coc{merge_Branching_SNNS}. If two branching terms have overlapping branches, then we try to merge them as exemplified by rule \coc{merge_Branching_SSSS}.\footnote{Due to space constraints, the names of these rules have been abbreviated in \cref{fig:merge}.}

\begin{figure*}
  \begin{eqnarray*}
    &\infer[\mbox{\lstinline+merge_End+}]
          {\mbox{\lstinline+End [V] End == End+}}
          {}
    \qquad
    \infer[\mbox{\lstinline+merge_Sel+}]
          {\mbox{\lstinline"p+l; B1 [V] p+l; B2 == p+l; B"}}
          {\mbox{\lstinline+B1 [V] B2 == B+}}
    \rulebreak
    &\infer[\mbox{\lstinline+merge_Cond+}]
          {\mbox{\lstinline+If p.e Then Bt1 Else Bt2 [V] If p.e Then Be1 Else Be2 == If p Then Bt Else Be+}}
          {\mbox{\lstinline+Bt1 [V] Bt2 == Bt+}
            & \mbox{\lstinline+Be1 [V] Be2 == Be+}}
    \rulebreak
    &\infer[\mbox{\lstinline+NNNN+}]
          {\mbox{\lstinline+p & None // None [V] p & None // None == p & None // None+}}
          {}
    \rulebreak
    &\infer[\mbox{\lstinline+SNNN+}]
          {\mbox{\lstinline+p & Some bL // None [V] p & None // None == p & Some bL // None+}}
          {}
    \rulebreak
    &\infer[\mbox{\lstinline+SNNS+}]
          {\mbox{\lstinline+p & Some bL // None [V] p & None // Some bR == p & Some bL // Some bR+}}
          {}
    \rulebreak
    &\infer[\mbox{\lstinline+SNSN+}]
          {\mbox{\lstinline+p & Some bL1 // None [V] p & Some bL2 // None == p & Some bL // None+}}
          {\mbox{\lstinline+bL1 [V] bL2 == bL+}}
    \rulebreak
    &\infer[\mbox{\lstinline+SSSN+}]
          {\mbox{\lstinline+p & Some bL1 // Some bR [V] p & Some bL2 // None == p & Some bL // Some bR+}}
          {\mbox{\lstinline+bL1 [V] bL2 == bL+}}
    \rulebreak
    &\infer[\mbox{\lstinline+SSSS+}]
          {\mbox{\lstinline+p & Some bL1 // Some bR1 [V] p & Some bL2 // Some bR2 == p & Some bL // Some bR+}}
          {\mbox{\lstinline+bL1 [V] bL2 == bL+}
            & \mbox{\lstinline+bR1 [V] bR2 == bR+}}
  \end{eqnarray*}
  \caption{Definition of the merge relation (selected rules).}
  \label{fig:merge}
\end{figure*}
As we remarked, merging (seen as a partial function) can be undefined, for example \coc{End} and \coc{p+l; End} cannot be merged.
This gives rise to the notion of \emph{projectability} anticipated in \cref{sec:intro}: a choreography \coc{C} is projectable on a process \coc{p} in the context of a set of procedure definitions \coc{D} if \coc{bproj} is defined for those parameters.
\begin{lstlisting}
Definition projectable_B D C p := exists B, [[D,C | p]] == B.
\end{lstlisting}
This is generalised by \coc{projectable_C D C ps}, which states that \coc{C} is projectable for all processes in the list \coc{ps}.
For a choreographic program \coc{P} to be projectable, written \coc{projectable_P P}, we require that \coc{Main P} be projectable for all processes in \coc{CCP_pn P} and that all procedures be projectable for the processes that they use.

With projectability in place, Endpoint Projection (EPP) is defined as a function that maps a projectable choreographic program to a process program in SP.
\begin{lstlisting}
Definition epp P : projectable_P P -> Program.
\end{lstlisting}
The second argument of \coc{epp} is a proof of \coc{projectable_P P}, but it is shown that the result does not depend on this term.

\begin{example}
The behaviours of \coc{buyer} and \coc{seller} in \cref{ex:bs-epp} are the respective projections for these two processes of the choreography in \cref{eq:bs-projectable}.
\end{example}

\subsection{Turing completeness}
The authors of~\cite{CMP21a} formalise that CC is Turing-complete, in the sense that all of Kleene's partial recursive functions~\cite{Kleene52} can be implemented as a choreography for a suitable notion of implementation.
The proof is interesting because it considers CC instantiated with very restricted computational capabilities at processes: values are natural numbers; expressions can only be the constant zero, a variable, or the successor of a variable; and Boolean expressions can only check if the two variables at a process contain the same value. Kleene's partial recursive functions are then implemented concurrently, by making processes communicate according to appropriate patterns.

A choreographic program \coc{P} implements \coc{f:PRFunction m} (representing a partial recursive function $f:\NN^m\to\NN$) with
input processes \coc{ps$_1$},\ldots,\coc{ps$_{m}$} and output process \coc{q} iff: for any state \coc{s} where
\coc{ps$_{1}$},\ldots,\coc{ps$_{m}$} contain the values \coc{n$_{1}$},\ldots,\coc{n$_{m}$} in their variable \coc{x}, (i) if
$f(\coc{n}_1,\ldots,\coc{n}_m) = \coc{n}$, then all executions of \coc{P} from \coc{s} terminate, and do so in a state where \coc{q}
stores \coc{n} in its variable \lstinline+x+; and (ii) if $f(\coc{n}_1,\ldots,\coc{n}_m)$ is undefined, then
execution of \coc{P} from \coc{s} never terminates.\footnote{This is a straightforward adaption of the definition of function implementation by a Turing machine~\cite{Turing36}.}
This is captured by the Coq term \coc{implements P m f ps q}, where \coc{ps} is the vector \coc{ps$_{1}$},\ldots,\coc{ps$_{m}$}.

The proof of Turing completeness encodes partial recursive functions to choreographies that are not always projectable, since they contain no selections but some processes behave differently in conditionals.

\section{Amendment}
\label{sec:amend}

Several works have studied how unprojectable choreographies can be automatically amended to obtain projectable versions~\cite{BB16,CM20,LMZ13}.
In particular, \cite{CM20} developed an amendment procedure based on merging.
The idea is that, whenever a choreography contains a conditional, amendment adds selections, in both branches, from the process evaluating the guard to any processes whose behaviour projection is undefined.
Intuitively, this makes the output choreography projectable.

\begin{example}\label{ex:amend-fail}
Let \coc{C} be the choreography:
\begin{lstlisting}
p.e --> q.x; If r.b Then (r.e' --> p.y; End) Else End
\end{lstlisting}
Amending \coc{C} as described yields the following choreography, \coc{A}:
\begin{lstlisting}
p.e --> q.x; If r.b Then (r --> p[l]; r.e' --> p.y; End)
                   Else (r --> p[r]; End)
\end{lstlisting}
\end{example}

Amendment is claimed to have the following properties.
\begin{lemma}[Amendment Lemma~\cite{CM20}, rephrased]\label{lem:amend-wrong}
For every choreography \coc{C}:
\begin{enumerate}
\item The amendment of \coc{C} is well-formed.
\item The amendment of \coc{C} is projectable.
\item If \coc{DA}, \coc{A}, and \coc{A'} are obtained by amending all procedures in \coc{D} as well as \coc{C} and \coc{C'}, then \coc{(D,C,s) --[tl]-->** (D,C',s')} iff \coc{(DA,A,s) --[tl']-->** (DA,A',s')} for some \coc{tl'}.
\end{enumerate}
\end{lemma}
In point one, well-formedness refers to a set of syntactic conditions that exclude ill-written choreographies, e.g., self-communications (interactions where a process communicates with itself)~\cite{CM20}.
Points one and two are simple to prove by induction on the structure of the choreography.
Point three, unfortunately, is wrong.
When attempting to formalise this result, we failed, and the state of the proof led us to the following counterexample.
\begin{example}
Given a suitable state, the choreography \coc{C} from \cref{ex:amend-fail} can make a transition to \coc{C'} defined as
\begin{lstlisting}
p.e --> q.x; r.e' --> p.y; End
\end{lstlisting}
by rules \coc{CC_Delay_Eta} and \coc{CC_Then}.
However, \coc{C}'s amendment \coc{A} can move to
\begin{lstlisting}
p.e --> q.x; r --> p[l]; r.e' --> p.y; End
\end{lstlisting}
by the same rules, but this is neither the amendment of \coc{C'}, nor can it reach it since the offending selection term is blocked by the initial communication.
\end{example}

In hindsight, this is not so surprising: amendment introduces causal dependencies that were not present in the source choreography.
However, this intuition was completely missed by both authors and reviewers of the original publications discussing amendment~\cite{CM16a,CM20}.
Therefore, amending a choreography can remove some execution paths.

In the rest of this section, we show how to define amendment formally in Coq, and formulate a correct variation of \cref{lem:amend-wrong}.

\subsection{Definition}
We decompose the definition of amendment in three functions: one for identifying the processes that need to be informed of the outcome of a specific conditional; one for prepending a list of selections to a choreography; and one that recursively amends a whole choreography by using the former two.
This division simplifies not only the definition, but also the structure of proofs about amendment since they can be modularised.

To identify the processes that require knowledge of choice, we define a function \coc{up_list} (\coc{up} is short for ``unprojectable processes'').
This function recursively goes through a list \coc{ps} of processes and checks for each process in the list whether the choreography \coc{If p.b Then C1 Else C2} can be projected on that process (function \coc{projectable_B_dec} does precisely this test).
If this is not the case, then the process is added to the result.
(Since projectability is relative to a set of procedure definitions, this also needs to be given as an argument, \coc{D}.)
\begin{lstlisting}
Fixpoint up_list D p b ps C1 C2 : list Pid := match ps with
  | nil => nil
  | r :: ps' => let ps'' := up_list D p b ps' C1 C2 in
    if (r =? p) then ps''
    else if projectable_B_dec D (If p.b Then C1 Else C2) r
         then ps''
         else (r :: ps'') end.
\end{lstlisting}
Note that \coc{p}, as the evaluator of the conditional, does not need to be informed of the outcome. This justifies the check \coc{r =? p}, whose inclusion also avoids introducing self-communications and simplifies subsequent proofs.

The second ingredient is straightforward: given a process \coc{p}, a selection label \coc{l}, and a choreography \coc{C}, it recursively adds selections of \coc{l} from \coc{p} to each element of a list \coc{ps}.
\begin{lstlisting}
Fixpoint add_sels p l ps C : Choreography := match ps with
  | nil => C
  | r :: ps' => p --> r[l]; add_sels p l ps' C end.
\end{lstlisting}

We can now define amendment following the informal procedure described in~\cite{CM20}. Given a list of processes \coc{ps}, we go through a choreography \coc{C}; whenever we meet a conditional on a process \coc{p}, we compute the list of processes from \coc{ps} with an undefined projection and prepend the branches of the conditional with appropriate selections. (We show only the most interesting cases.)
\begin{lstlisting}
Fixpoint amend D ps C := match C with
  | eta; C' => eta; (amend D ps C')
  | If p.b Then C1 Else C2 =>
      let l := up_list D p b ps (amend D ps C1) (amend D ps C2) in
      If p.b Then (add_sels p left l (amend D ps C1))
             Else (add_sels p right l (amend D ps C2))
  | ... end.
\end{lstlisting}

Amendment is generalised to sets of procedure definitions in the obvious way.
\begin{lstlisting}
Definition amend_D D ps : DefSet := fun X => (fst (D X), amend D ps (snd (D X))).
\end{lstlisting}

To amend a program \coc{P}, the parameter \coc{ps} of the previous functions is instantiated with the set of processes used in \coc{P}.
\begin{lstlisting}
Definition amend_P P :=
  (amend_D (Procedures P) (CCP_pn P), amend (Procedures P) (CCP_pn P) (Main P)).
\end{lstlisting}
This formal definition corresponds to the informal one given in~\cite{CM20}.
In particular, all our examples are formalised in Coq.

\begin{example}
Consider the following choreography.
\begin{lstlisting}
If p.b Then (p.e --> q.x; q.e' --> r.y; End)
       Else (q.e'' --> r.y; End)
\end{lstlisting}
Here, \coc{p} decides if (i) it will communicate a value to \coc{q} that can be used in the computation of a later message from \coc{q} to \coc{r} (so \coc{q} acts as a sort of proxy) or (ii) \coc{q} should just compute the value that it will communicate to \coc{r} by itself.
Amendment is smart enough to notice that while \coc{q} requires a selection from \coc{p}, \coc{r} does not since it behaves in the same way (receive from \coc{q} on \coc{x}). Therefore, amending the choreography returns the following.
\begin{lstlisting}
If p.b Then (p --> q[left]; p.e --> q.x; q.e' --> r.y; End)
       Else (p --> q[right]; q.e'' --> r.y; End)
\end{lstlisting}
\end{example}

\subsection{Syntactic Properties}
\label{sec:amend-syntax}
We now discuss the key properties of amendment.

Amendment preserves well-formedness of choreographies (\coc{Choreography_WF}) and choreographic programs (\coc{Program_WF}). This follows from the fact that \coc{add_sels} preserves all syntactic properties of well-formedness, using induction.
\begin{lstlisting}
Lemma amend_Choreography_WF : Choreography_WF C -> Choreography_WF (amend D ps C).

Lemma amend_Program_WF : Program_WF (D,C) -> Program_WF (amend_D D ps,amend D ps C).
\end{lstlisting}
(For simplicity, we omit universal quantifiers at the beginning of lemmas.)

Likewise, it is straightforward to prove that amending for some processes guarantees that the output choreography is projectable on all those processes.
\begin{lstlisting}
Lemma amend_projectable_C : projectable_C (amend_D D ps) (amend D ps C) ps.
\end{lstlisting}
We do not generalise this result to choreographic programs: it is not straightforward to do and our later development does not need it.
The issue we encounter is related to a problem discussed in~\cite{CMP21a,CMP21b}: computing the set of processes and procedures that are used by a choreography can require an infinite number of steps, and is therefore not definable as a function in Coq.
(A simple example is a program with an infinite set of procedure definitions where each procedure \coc{X}$_i$ invokes the next procedure \coc{X}$_{i+1}$.)

The function \coc{CCP_pn} used in the definition of \coc{amend_P} does return the set of processes involved in a program \coc{P}, but it does not check that \coc{P} does not define unused procedures.
If this is the case, these procedures may use processes not in \coc{CCP_pn P}, and therefore they may be unprojectable for these processes.
Rather than stating a result with complex side-conditions as hypotheses, we prove projectability of particular programs applying \coc{amend_projectable_C} to \coc{Main P} and to the bodies of all procedure definitions.
The development in the next section uses this strategy.

\subsection{Semantic Properties}
We now discuss how the formulation of the semantic relation between a choreography and its amendment needs to be changed.

The counterexample shown earlier suggests allowing both choreographies to perform additional transitions in order to unblock and remove lingering selections introduced by amendment.
(In our example, this would be the communication from \coc{p} to \coc{q}.)
The correspondence would then look as follows, where the dotted lines correspond to existentially quantified terms:
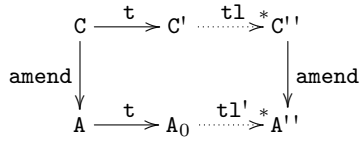
\[
\xymatrix{
  \coc{C} \ar[r]^{\coc{t}} \ar[d]_{\coc{amend}}
  & \coc{C'} \ar@{.>}[r]^{\coc{tl}}^(0.78){*}
  & \coc{C''} \ar[d]^{\coc{amend}}
  \\
  \coc{A} \ar[r]^{\coc{t}}
  & \coc{A}_0 \ar@{.>}[r]^{\coc{tl'}}^(0.78){*}
  & \coc{A''}
}
\]
and the list of transition labels \coc{tl} can be obtained from \coc{tl'} by removing some selections.

Our attempt to prove this result showed that it holds for all cases but one: when the transition \coc{t} is obtained by applying rule \coc{CC_Delay_Cond}.
\begin{example}
We show a minimal counterexample.
Consider the choreography
\begin{lstlisting}
If p.b Then (q.e --> r.x; q.e --> p.x; End)
       Else (q.e --> r.x; End)
\end{lstlisting}
and its amendment
\begin{lstlisting}
If p.b Then (p --> q[left]; q.e --> r.x; q.e --> p.x; End)
       Else (p --> q[right]; q.e --> r.x; End) .
\end{lstlisting}
The original choreography can execute the communication between \coc{q} and \coc{r}, reaching
\begin{lstlisting}
If p.e Then (q.e' --> p.x; End) Else End
\end{lstlisting}
but its amendment needs to run the conditional and a selection before it can execute the same communication.
\end{example}

There are two ways to solve this problem: changing the definition of amendment, or refining the correspondence result further.
We opted for the second route, for two reasons: first, we get to keep the original definition given on paper in~\cite{CM20}; second, making amendment clever enough to recognise this kind of situations requires a non-local analysis of the choreography (i.e., looking at the structure of the branches of conditionals instead of simply checking for projectability of the term). In our example, such an analysis could detect that the additional selections from \coc{p} to \coc{q} could be added only after the communication from \coc{q} to \coc{r}, solving the issue.

Therefore, our final correspondence result requires that the amendment of a choreography be allowed to perform additional transitions \emph{before} it matches the transition performed by the original choreography.
Since a transition may invoke rule \coc{c_delay_Cond} more than once, this means that the orders of the transitions performed by the original choreography and its amendment can be arbitrary permutations of each other (ignoring the extra selections).

The correspondence result we prove looks as follows:
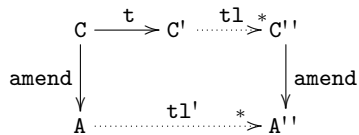
\[
\xymatrix{
  \coc{C} \ar[r]^{\coc{t}} \ar[d]_{\coc{amend}}
  & \coc{C'} \ar@{.>}[r]^{\coc{tl}}^(0.78){*}
  & \coc{C''} \ar[d]^{\coc{amend}}
  \\
  \coc{A} \ar@{.>}[rr]^{\coc{tl'}}^(0.78){*}
  &
  & \coc{A''}
}
\]
where \coc{t::tl} can be obtained from \coc{tl'} by removing some selections and permuting labels.

To formalise this in Coq, we introduce a relation \coc{sel_exp} (``selection expansion'') between lists of transition labels.
\begin{lstlisting}
Inductive sel_exp :=
| se_base tl tl' : Permutation tl tl' -> sel_exp tl tl'
| se_extra p q l tl tl' tl'' : sel_exp tl tl' ->
    Permutation (TL_Sel p q l::tl') tl'' -> sel_exp tl tl''.
\end{lstlisting}

We can now prove a correct version of the correspondence between choreographies and their amendments. There are four results in total: the one depicted above and its generalisation to the case where \coc{t} is replaced with a list of transition labels; and the two dual results where the amendment of a choreography moves first. We show the two more general statements.
\begin{lstlisting}
Lemma amend_complete_many : Program_WF (D,C) -> (D,C,s) --[tl]-->** (D,C',s') ->
  exists tl' tl'' C'' s'', sel_exp (tl++tl') tl''  /\ (D,C',s') --[tl']-->** (D,C'',s'')
    /\ (amend_D D ps,amend D ps C,s) --[tl'']-->** (amend_D D ps,amend D ps C'',s'').
  
Lemma amend_sound_many : Program_WF (D,C) -> let (D' := amend_D D ps) in
  (D', amend D ps C,s) --[tl]-->** (D',C',s') ->
  exists tl' tl'' C'' s'', (D',C',s') --[tl']-->** (D', amend D ps C'',s'')
    /\ (D,C,s) --[tl'']-->** (D,C'',s'') /\ sel_exp tl'' (tl++tl').
\end{lstlisting}

The challenging part of the work in this section was understanding what the correct formulation of these results should be. Once we reached this formulation, proofs were relatively straightforward inductions on the given transitions (10--15 lines of Coq code per case).

The formalisation of the amendment lemma consists of 6 definitions, 50 lemmas, and 4 examples, with a total of roughly 1050 lines of Coq code.

\section{Implications of Amendment}
\label{sec:turing}
In the previous section, we had to weaken the original statement for the semantic correspondence guaranteed by amendment that was given in~\cite{CM20}.
Since the original statement was used in the proofs of Turing completeness for projectable core choreographies and SP, it is natural to investigate whether our new formulation still yields these results.

For uniformity, we start by reformulating the Turing completeness result for core choreographies from~\cite{CMP21a}, where process names are identified with natural numbers.
\begin{lstlisting}
Theorem CC_Turing_Complete : forall n (f:PRFunction n),
  exists P, Program_WF P /\ implements P f (vec_1_to_n n) 0.
\end{lstlisting}
The theorem states that, for any partial recursive function \coc{f}, there exists a well-formed choreographic program \coc{P} that implements \coc{f} with input processes \coc{1}, \ldots, \coc{n} and output process \coc{0}.
The proof is a straightforward combination of results already presented in~\cite{CMP21a}.

Combining this result with our lemmas about amendment yields that the fragment of projectable core choreographies is also Turing-complete.
\begin{lstlisting}
Lemma projCC_Turing_Complete : forall n (f:PRFunction n),
  exists P, Program_WF P /\ projectable_P P /\ implements P f (vec_1_to_n n) 0.
\end{lstlisting}
The proof is split into several steps. The most interesting sublemma is the one establishing that amending a choreography that implements a function yields a choreography that implements the same function. This is formulated as a general result about amendment.
\begin{lstlisting}
Lemma amend_implements : Program_WF P ->
  implements P f ps q -> implements (amend_P P) f ps q.
\end{lstlisting}
The proof uses the fact that terminated choreographies cannot execute further to show that the list of additional transitions added to the original choreography by the amendment lemma (\coc{tl} in the last diagram) must be empty.

The remaining lemmas for \coc{projCC_Turing_Complete} deal with projectability of the amended choreography, as discussed in the previous section, and are simple to prove.

Since amended choreographies are projectable, we can further apply the EPP theorem from~\cite{CMP21b} to show that SP is also Turing-complete.
\begin{lstlisting}
Theorem SP_Turing_Complete : forall n (f:PRFunction n),
  exists P, Network_WF (Net P) /\ SP_implements P f (vec_1_to_n n) 0.
\end{lstlisting}
The definition of \coc{SP_implements} is a straightforward adaptation of the definition of \coc{implements} for choreographies.
The proof of \coc{SP_Turing_Complete} follows a similar strategy to the one for \coc{projCC_Turing_Complete}: we prove a sublemma \coc{epp_implements} stating that the EPP of a choreography that implements a function \coc{f} is a process program that implements \coc{f}.

The formalisation of this section consists of 2 definitions and 11 lemmas, totaling about 250 lines of Coq code.
The conciseness of this development substantiates our previous comment on not providing a complex lemma for projectability of programs, at the end of \cref{sec:amend-syntax}.

\section{Related Work}
\label{sec:related}

To the best of our knowledge, our work is the first formalisation of amendment, its properties, and its intended consequences.

The work nearest to ours is the original presentation of the amendment procedure that inspired us~\cite{CM20}. As we discussed, the behavioural correspondence for amendment that the authors state is wrong. We developed a correct statement and managed to update and formalise the proofs of Turing completeness for CC and SP accordingly.
Our formalisation of the behavioural correspondence also clarifies what semantic property amendment actually guarantees, which might be important for future work and practical applications of amendment.

Amendment or similar procedures have been investigated also for other choreographic languages~\cite{BB16,LMZ13}. In all these works, the general idea is to make choreographies projectable by adding communications as needed. However, the differences between the underlying languages make the procedures very different from ours, which is based on merging. Merging was first introduced in~\cite{CHY12}.

Our work is based on the most recent version of the formalisation of CC, SP, and EPP~\cite{CMP22}, which was originally introduced in~\cite{CMP21b,CMP21a}. We did not need to modify this formalisation in order to use it for our development, which shows that it reached a sufficient level of maturity for being used as a library to reason about choreographies.

Other formalisations of choreographies include: Kalas, a choreographic programming language that targets CakeML~\cite{PGSN22}; the choreographic DSL Zooid, a Coq library for verifying that message passing code respects a given multiparty session type (these are abstract choreographies without computation)~\cite{CFGY21}; and multiparty GV, a formalised functional language with a similar goal to Zooid~\cite{JBK22}.

\section{Conclusion}
\label{sec:conclusion}

We have presented the first formalisation of an amendment procedure for choreographies.
Our work is based on a previous formalisation of CC and its accompanying notion of EPP, which we used as a library.
We found this formalisation to be modular and complete enough to support the separate development presented here.
In the same spirit of generality and reusability, our formalisation does not add any assumptions about CC that were not present in the library.

Our development is an illustration of how theorem provers can assist in research: interacting with Coq guided us to (i) discovering that the semantic property of amendment found in the background literature for this work is wrong, and (ii) a correct formulation that is still powerful enough for its intended use in previous work.

The formalisation of amendment is amenable to extraction, and therefore our work offers a basis for a certified transformer from arbitrary choreographies in CC to projectable ones.
In the future, we plan on studying how this transformer can be integrated into existing frameworks for choreographic programming.

Our notion of amendment is intrinsically related to how EPP is defined for CC. In the literature, there are choreographic languages with a more permissive notion of knowledge of choice, e.g., where replicated processes intended to be used as services are allowed to be involved in only one branch of a conditional~\cite{CHY12,CM13}. It would be interesting to study how amendment can be adapted to these settings.

\bibliography{biblio}

\begin{thebibliography}{10}

\bibitem{BB16}
Samik Basu and Tevfik Bultan.
\newblock Automated choreography repair.
\newblock In Perdita Stevens and Andrzej Wasowski, editors, {\em Procs.\ FASE},
  volume 9633 of {\em Lecture Notes in Computer Science}, pages 13--30.
  Springer, 2016.
\newblock \href {https://doi.org/10.1007/978-3-662-49665-7_2}
  {\path{doi:10.1007/978-3-662-49665-7_2}}.

\bibitem{CHY12}
Marco Carbone, Kohei Honda, and Nobuko Yoshida.
\newblock Structured communication-centered programming for web services.
\newblock {\em {ACM} Trans.\ Program.\ Lang.\ Syst.}, 34(2):8:1--8:78, 2012.
\newblock \href {https://doi.org/10.1145/2220365.2220367}
  {\path{doi:10.1145/2220365.2220367}}.

\bibitem{CLMSW16}
Marco Carbone, Sam Lindley, Fabrizio Montesi, Carsten Sch{\"{u}}rmann, and
  Philip Wadler.
\newblock Coherence generalises duality: {A} logical explanation of multiparty
  session types.
\newblock In Jos{\'{e}}e Desharnais and Radha Jagadeesan, editors, {\em Procs.\
  CONCUR}, volume~59 of {\em LIPIcs}, pages 33:1--33:15. Schloss Dagstuhl -
  Leibniz-Zentrum f{\"{u}}r Informatik, 2016.

\bibitem{CM13}
Marco Carbone and Fabrizio Montesi.
\newblock Deadlock-freedom-by-design: multiparty asynchronous global
  programming.
\newblock In Roberto Giacobazzi and Radhia Cousot, editors, {\em Procs.\ POPL},
  pages 263--274. {ACM}, 2013.
\newblock \href {https://doi.org/10.1145/2429069.2429101}
  {\path{doi:10.1145/2429069.2429101}}.

\bibitem{CDP11}
Giuseppe Castagna, Mariangiola Dezani{-}Ciancaglini, and Luca Padovani.
\newblock On global types and multi-party sessions.
\newblock In Roberto Bruni and J{\"{u}}rgen Dingel, editors, {\em Formal
  Techniques for Distributed Systems - Joint 13th {IFIP} {WG} 6.1 International
  Conference, {FMOODS} 2011, and 31st {IFIP} {WG} 6.1 International Conference,
  {FORTE} 2011, Reykjavik, Iceland, June 6-9, 2011. Proceedings}, volume 6722
  of {\em Lecture Notes in Computer Science}, pages 1--28. Springer, 2011.
\newblock \href {https://doi.org/10.1007/978-3-642-21461-5_1}
  {\path{doi:10.1007/978-3-642-21461-5_1}}.

\bibitem{CFGY21}
David Castro{-}Perez, Francisco Ferreira, Lorenzo Gheri, and Nobuko Yoshida.
\newblock Zooid: a {DSL} for certified multiparty computation: from mechanised
  metatheory to certified multiparty processes.
\newblock In Stephen~N. Freund and Eran Yahav, editors, {\em Procs.\ PLDI},
  pages 237--251. {ACM}, 2021.
\newblock \href {https://doi.org/10.1145/3453483.3454041}
  {\path{doi:10.1145/3453483.3454041}}.

\bibitem{CM16a}
Lu{\'{\i}}s Cruz{-}Filipe and Fabrizio Montesi.
\newblock A core model for choreographic programming.
\newblock In Olga Kouchnarenko and Ramtin Khosravi, editors, {\em Procs.\
  FACS}, volume 10231 of {\em Lecture Notes in Computer Science}, pages 17--35.
  Springer, 2017.
\newblock \href {https://doi.org/10.1007/978-3-319-57666-4_3}
  {\path{doi:10.1007/978-3-319-57666-4_3}}.

\bibitem{CM20}
Lu{\'{\i}}s Cruz{-}Filipe and Fabrizio Montesi.
\newblock A core model for choreographic programming.
\newblock {\em Theor.\ Comput.\ Sci.}, 802:38--66, 2020.
\newblock \href {https://doi.org/10.1016/j.tcs.2019.07.005}
  {\path{doi:10.1016/j.tcs.2019.07.005}}.

\bibitem{CMP21b}
Lu{\'{\i}}s Cruz{-}Filipe, Fabrizio Montesi, and Marco Peressotti.
\newblock Certifying choreography compilation.
\newblock In Antonio Cerone and Peter~Csaba {\"{O}}lveczky, editors, {\em
  Procs.\ ICTAC}, volume 12819 of {\em LNCS}, pages 115--133. Springer, 2021.
\newblock \href {https://doi.org/10.1007/978-3-030-85315-0_8}
  {\path{doi:10.1007/978-3-030-85315-0_8}}.

\bibitem{CMP21a}
Lu{\'{\i}}s Cruz{-}Filipe, Fabrizio Montesi, and Marco Peressotti.
\newblock Formalising a {Turing-}complete choreographic language in {Coq}.
\newblock In Liron Cohen and Cezary Kaliszyk, editors, {\em Procs.\ ITP},
  volume 193 of {\em LIPIcs}, pages 15:1--15:18. Schloss Dagstuhl --
  Leibniz-Zentrum f{\"{u}}r Informatik, 2021.
\newblock \href {https://doi.org/10.4230/LIPIcs.ITP.2021.15}
  {\path{doi:10.4230/LIPIcs.ITP.2021.15}}.

\bibitem{CMP22}
Luís Cruz-Filipe, Fabrizio Montesi, and Marco Peressotti.
\newblock A formal theory of choreographic programming.
\newblock {\em CoRR}, abs/2209.01886, 2022.
\newblock URL: \url{https://arxiv.org/abs/2209.01886}.

\bibitem{DGGLM17}
Mila Dalla~Preda, Maurizio Gabbrielli, Saverio Giallorenzo, Ivan Lanese, and
  Jacopo Mauro.
\newblock Dynamic choreographies: Theory and implementation.
\newblock {\em Log.\ Methods Comput.\ Sci.}, 13(2), 2017.
\newblock \href {https://doi.org/10.23638/LMCS-13(2:1)2017}
  {\path{doi:10.23638/LMCS-13(2:1)2017}}.

\bibitem{FL17}
Alain Finkel and {\'{E}}tienne Lozes.
\newblock Synchronizability of communicating finite state machines is not
  decidable.
\newblock In Ioannis Chatzigiannakis, Piotr Indyk, Fabian Kuhn, and Anca
  Muscholl, editors, {\em 44th International Colloquium on Automata, Languages,
  and Programming, {ICALP} 2017, July 10-14, 2017, Warsaw, Poland}, volume~80
  of {\em LIPIcs}, pages 122:1--122:14. Schloss Dagstuhl - Leibniz-Zentrum
  f{\"{u}}r Informatik, 2017.
\newblock \href {https://doi.org/10.4230/LIPIcs.ICALP.2017.122}
  {\path{doi:10.4230/LIPIcs.ICALP.2017.122}}.

\bibitem{GMP20}
Saverio Giallorenzo, Fabrizio Montesi, and Marco Peressotti.
\newblock Choreographies as objects.
\newblock {\em CoRR}, abs/2005.09520, 2020.
\newblock URL: \url{https://arxiv.org/abs/2005.09520}.

\bibitem{HG22}
Andrew~K. Hirsch and Deepak Garg.
\newblock Pirouette: higher-order typed functional choreographies.
\newblock {\em Proc. {ACM} Program. Lang.}, 6({POPL}):1--27, 2022.
\newblock \href {https://doi.org/10.1145/3498684} {\path{doi:10.1145/3498684}}.

\bibitem{HYC16}
Kohei Honda, Nobuko Yoshida, and Marco Carbone.
\newblock Multiparty asynchronous session types.
\newblock {\em J. {ACM}}, 63(1):9, 2016.
\newblock Also: POPL, pages 273--284, 2008.
\newblock \href {https://doi.org/10.1145/2827695} {\path{doi:10.1145/2827695}}.

\bibitem{JBK22}
Jules Jacobs, Stephanie Balzer, and Robbert Krebbers.
\newblock Multiparty gv: Functional multiparty session types with certified
  deadlock freedom.
\newblock Accepted for publication at ICFP 2022, 2022.

\bibitem{JB22}
Sung{-}Shik Jongmans and Petra van~den Bos.
\newblock A predicate transformer for choreographies - computing preconditions
  in choreographic programming.
\newblock In Ilya Sergey, editor, {\em Programming Languages and Systems - 31st
  European Symposium on Programming, {ESOP} 2022, Held as Part of the European
  Joint Conferences on Theory and Practice of Software, {ETAPS} 2022, Munich,
  Germany, April 2-7, 2022, Proceedings}, volume 13240 of {\em Lecture Notes in
  Computer Science}, pages 520--547. Springer, 2022.
\newblock \href {https://doi.org/10.1007/978-3-030-99336-8_19}
  {\path{doi:10.1007/978-3-030-99336-8_19}}.

\bibitem{Kleene52}
Stephen~Cole Kleene.
\newblock {\em Introduction to Metamathematics}, volume~1.
\newblock North-Holland Publishing Co., 1952.

\bibitem{LMZ13}
Ivan Lanese, Fabrizio Montesi, and Gianluigi Zavattaro.
\newblock Amending choreographies.
\newblock In Ant{\'{o}}nio Ravara and Josep Silva, editors, {\em Proceedings
  9th International Workshop on Automated Specification and Verification of Web
  Systems, {WWV} 2013, Florence, Italy, 6th June 2013}, volume 123 of {\em
  {EPTCS}}, pages 34--48, 2013.
\newblock \href {https://doi.org/10.4204/EPTCS.123.5}
  {\path{doi:10.4204/EPTCS.123.5}}.

\bibitem{LLLG16}
Tanakorn Leesatapornwongsa, Jeffrey~F. Lukman, Shan Lu, and Haryadi~S. Gunawi.
\newblock Taxdc: {A} taxonomy of non-deterministic concurrency bugs in
  datacenter distributed systems.
\newblock In Tom Conte and Yuanyuan Zhou, editors, {\em Proceedings of the
  Twenty-First International Conference on Architectural Support for
  Programming Languages and Operating Systems, {ASPLOS} 2016, Atlanta, GA, USA,
  April 2-6, 2016}, pages 517--530. {ACM}, 2016.
\newblock \href {https://doi.org/10.1145/2872362.2872374}
  {\path{doi:10.1145/2872362.2872374}}.

\bibitem{LN15}
Alberto Lluch{-}Lafuente, Flemming Nielson, and Hanne~Riis Nielson.
\newblock Discretionary information flow control for interaction-oriented
  specifications.
\newblock In Narciso Mart{\'{\i}}{-}Oliet, Peter~Csaba {\"{O}}lveczky, and
  Carolyn~L. Talcott, editors, {\em Logic, Rewriting, and Concurrency}, volume
  9200 of {\em Lecture Notes in Computer Science}, pages 427--450. Springer,
  2015.
\newblock \href {https://doi.org/10.1007/978-3-319-23165-5_20}
  {\path{doi:10.1007/978-3-319-23165-5_20}}.

\bibitem{LH17}
Hugo~A. L{\'{o}}pez and Kai Heussen.
\newblock Choreographing cyber-physical distributed control systems for the
  energy sector.
\newblock In Ahmed Seffah, Birgit Penzenstadler, Carina Alves, and Xin Peng,
  editors, {\em Procs.\ SAC}, pages 437--443. {ACM}, 2017.
\newblock \href {https://doi.org/10.1145/3019612.3019656}
  {\path{doi:10.1145/3019612.3019656}}.

\bibitem{M13p}
Fabrizio Montesi.
\newblock {\em Choreographic Programming}.
\newblock {Ph.{D}. Thesis}, IT University of Copenhagen, 2013.
\newblock URL:
  \url{http://www.fabriziomontesi.com/files/choreographic\_programming.pdf}.

\bibitem{M22}
Fabrizio Montesi.
\newblock {\em Introduction to Choreographies}.
\newblock Cambridge University Press, 2023.

\bibitem{NS78}
Roger~M. Needham and Michael~D. Schroeder.
\newblock Using encryption for authentication in large networks of computers.
\newblock {\em Commun.\ {ACM}}, 21(12):993--999, 1978.
\newblock \href {https://doi.org/10.1145/359657.359659}
  {\path{doi:10.1145/359657.359659}}.

\bibitem{PGSN22}
Johannes~{\AA}man Pohjola, Alejandro G{\'{o}}mez{-}Londo{\~{n}}o, James Shaker,
  and Michael Norrish.
\newblock Kalas: {A} verified, end-to-end compiler for a choreographic
  language.
\newblock In June Andronick and Leonardo de~Moura, editors, {\em 13th
  International Conference on Interactive Theorem Proving, {ITP} 2022, August
  7-10, 2022, Haifa, Israel}, volume 237 of {\em LIPIcs}, pages 27:1--27:18.
  Schloss Dagstuhl - Leibniz-Zentrum f{\"{u}}r Informatik, 2022.
\newblock \href {https://doi.org/10.4230/LIPIcs.ITP.2022.27}
  {\path{doi:10.4230/LIPIcs.ITP.2022.27}}.

\bibitem{Turing36}
Alan~M. Turing.
\newblock Computability and {$\lambda$}-definability.
\newblock {\em J. Symb.\ Log.}, 2(4):153--163, 1937.
\newblock \href {https://doi.org/10.2307/2268280} {\path{doi:10.2307/2268280}}.

\end{thebibliography}

\end{document}